\documentclass[12pt]{iopart}
\usepackage{graphicx}
\usepackage{bbm}

\begin{document}

\title{Theory of the strongly-damped quantum harmonic oscillator}

\author{Stephen M. Barnett}

\address{School of Physics and Astronomy, University of Glasgow, Glasgow G12 8QQ, UK}

\ead{stephen.barnett@glasgow.ac.uk} 

\author{James D. Cresser}

\address{Department of Physics and Astronomy, Faculty of Science and Engineering, Macquarie University,
NSW 2109, Australia}

\ead{james.cresser@mq.edu.au}

\author{Sarah Croke}

\address{School of Physics and Astronomy, University of Glasgow, Glasgow G12 8QQ, UK}

\ead{sarah.croke@glasgow.ac.uk} 

\date{\today}

\begin{abstract}
We analyse the properties of a strongly-damped quantum harmonic oscillator by means of an exact diagonalisation
of the full Hamiltonian, including both the oscillator and the reservoir degrees of freedom to which it is coupled. 
Many of the properties of the oscillator, including its steady-state properties and entanglement with the reservoir
can be understood and quantified in terms of a simple probability density, which we may associate with the ground-state
frequency spectrum of the oscillator.
\end{abstract}

\maketitle

\section{Introduction}
Recent technological advances make it possible to realise simple mechanical devices in the microscopic and 
nanoscopic regimes, the properties of which are determined by quantum effects \cite{Chan2011}.  The existence 
of these represents a remarkable opportunity for fundamental studies of light-matter interactions \cite{Gigan2006}
and also the potential for practical application to quantum communications and information processing
\cite{Stannigel2010}.  Yet they present also a challenge to existing methods of analysis, many of which were
developed to treat more rapidly oscillating systems with weaker couplings.  

The quantum theory of machines is built, to a large extent on the theory of oscillators and strongly coupled oscillators,
which are coupled to one or more environments each of which is at a characteristic temperature 
\cite{Zhang2014,Joshi,Brunelli2015}.  The behaviour of these quantum systems is governed not simply by
average properties but also fluctuations, and has been informed by the development of fluctuation theorems
for quantum open systems \cite{Campisi,Talkner}.  The coupling to the environment needs to be handled
with some care, however, because of the possibility of quantum coherence and the development of entanglement
between the system of interest and its environment, with the result that apparently unphysical behaviours may emerge
\cite{Allahverdyan}.  The requirement to treat the coupling to the reservoirs with care provides the incentive to return to 
the problem of a single strongly-damped oscillator and to treat this model exactly.

There exist at present mathematically and physically acceptable methods for treating strongly damped quantum 
oscillators, but most of these do not retain the simplicity of their classical counterparts.  Our aim is to recover at least some of 
this simplicity.  We do not seek to apply familiar techniques developed for weakly damped systems, but proceed instead 
by means of an exact diagonlisation of the oscillator-reservoir Hamiltonian.  This approach is complementary to that
adopted by Philbin who has tackled this problem of the oscillator evolution using Green functions \cite{Philbin2012}.  We
hope that the combination of his work and ours will provide a more complete understanding.

Our procedure produces 
a single positive function of frequency the integral of which is unity.  We can interpret this quantity as the frequency 
spectrum or probability density for the oscillator in its ground state.  The steady-state of the oscillator, its entanglement 
with the environment, its energy and much of its dynamics can be understood in terms of this probability.  Before
embarking on this programme, let us pause to review some of the background to our study.

\subsection{Background}

The harmonic oscillator has a special place in physics as one of the simplest and most widely employed of 
physical models.  The reasons for its ubiquity, no doubt, are its simplicity and the fact that it is readily analysed.
In the quantum domain, the harmonic oscillator is barely more difficult to treat than its classical counterpart and 
was one of the first dynamical systems to which Schr\"{o}dinger applied his equation \cite{Schrodinger}.  Today,
both the classical and quantum forms appear in elementary courses on classical and quantum mechanics.

The damped harmonic oscillator loses energy as a result of coupling to the surrounding environment.  In the
classical domain it often suffices to describe this in terms of a simple damping coefficient, $\gamma$, and an
associated stochastic or Langevin force \cite{Langevin}, $F(t)$, which models the effect of environmental 
fluctuations on the oscillator \cite{Uhlenbeck1930,Wang1945,Risken,Lemons,Mazo}.  The dynamics is described by 
a simple linear differential equation of the form
\begin{equation}
\label{Eq1}
\ddot{x} + \gamma\dot{x} + \omega^2_0x = \frac{F(t)}{m} ,
\end{equation}
where $m$ is the mass of the oscillating particle.  There is no requirement for detailed knowledge of the 
fluctuating force, which may be considered to have a very short correlation time with a magnitude determined
by the requirements of thermodynamic equilibrium.

The damped quantum harmonic oscillator requires that explicit account be taken of the quantum nature of the
environmental degrees of freedom \cite{Senitzky}, which are most simply described by an ensemble of 
harmonic oscillators \cite{Louisell1964}.  If the damping is very weak, so that $\gamma \ll \omega_0$, then 
we can neglect rapidly oscillating terms in the coupling between the oscillator and the environmental oscillators
by making the rotating wave approximation, which corresponds to enforcing the conservation of the total number
of vibrational quanta, and then the Born and Markov approximations
associated with weak coupling and loss of memory in the reservoir \cite{Radmore1997}.  This leads to the
master equations and Heisenberg-Langevin equations that are ubiquitous, most especially, in quantum optics
\cite{Radmore1997,Louisell1973,Perina,Meystre1991,Carmichael1993,Walls1994,Carmichael1999,Breuer2002,
Gardiner2004,Ficek2004,Wiseman2010,Weiss2012}.  

If the coupling is somewhat stronger then it may not be possible to make the rotating wave approximation and
we then need to retain in the Hamiltonian terms that can create or annihilate a pair of quanta,
one in the damped oscillator and, at the same time, one in the environment.  This leads to the Caldeira-Leggett
model \cite{Philbin2012,Agarwal,UllersmaI,UllersmaII,CalderiaI,CaldeiraII,Haake}, which we describe in the the 
following section, and which has been applied to study a wide variety of quantum open systems \cite{Breuer2002,Weiss2012}.  
A variant on the model has been applied to the quantum theory of light in dielectric \cite{Huttner1992a,Huttner1992b} 
and magneto-dielectric media \cite{Kheirandish2006,Kheirandish2008,Kheirandish2009,Amooshahi2009}.  An 
important complication that seems to be an inevitable consequence working in this strong-coupling regime is the 
failure of the Markov approximation; attempts to enforce this approximation lead to a master equation that is unphysical 
in that there exist initial states for which the dynamics leads to negative probabilities \cite{Breuer2002,Munro,Stenholm,Cresser2005,Cresser2006}.  It is possible to derive a master equation but 
the resulting equation is one that has within it non-trivial time-dependent coefficients \cite{Haake,Hu}.  This 
time-dependence is a clear signature of the non-Markovian nature of the associated evolution.  It seems that
this non-Markovian character is an inevitable feature of the strongly-damped quantum harmonic oscillator.


\section{Hamiltonian for the strongly-damped harmonic oscillator}

\begin{figure}[htbp] 
\centering
\includegraphics[width=10cm]{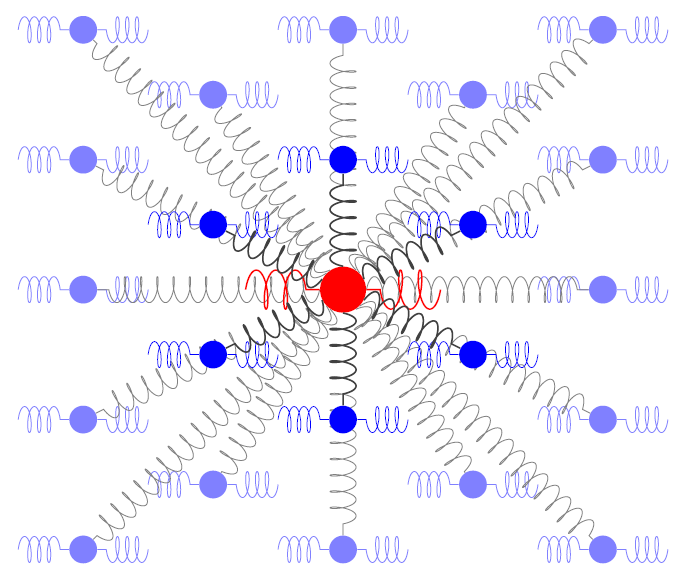}
\caption{Representation of a single harmonic oscillator (in red) coupled harmonically to a bath of 
oscillators (in blue).} 
\label{fig:figure1}
\end{figure}

Consider a harmonic oscillator of natural angular frequency $\omega_0$ that is strongly coupled to a
large collection of oscillators, with a range of frequencies, through their respective positions as depicted
in figure \ref{fig:figure1}.  We write
the Hamiltonian for the combined oscillator-reservoir system in the form \cite{CalderiaI,Haake}
\begin{equation}
\label{Eq2}
\hat{H} = \frac{\hat{p}^2}{2m} + \frac{1}{2}m\omega^2_0\hat{x}^2 + 
\sum_\mu\left(\frac{\hat{p}^2_\mu}{2m_\mu} + \frac{1}{2}m\omega^2_\mu\hat{x}_\mu\right)
- \sum_\mu m_\mu\omega_\mu^2\lambda_\mu\hat{x}_\mu\hat{x} .
\end{equation}
If we complete the square we can rewrite this in a minimal-coupling form to arrive at the alternative form:
\begin{equation}
\label{Eq3}
\hat{H} = \frac{\hat{p}^2}{2m} + \frac{1}{2}m\Omega^2_0\hat{x}^2 + 
\sum_\mu\left[\frac{\hat{p}^2_\mu}{2m_\mu} + 
\frac{1}{2}m_\mu\omega^2_\mu\left(\hat{x}_\mu - \lambda_\mu\hat{x}\right)^2\right] ,
\end{equation}
where 
\begin{equation}
\label{Eq4}
\Omega_0^2 = \omega_0^2 - \sum_\mu \frac{m_\mu}{m}\omega^2_\mu\lambda^2_\mu .
\end{equation}
Each term in our Hamiltonian is strictly positive only if this quantity is positive.  If it is negative then the
second term is also negative and the Hamiltonian is not bounded from below and hence not allowed
physically.  Hence the positivity of the Hamiltonian places a physical restriction on the strength of
the coupling
\begin{equation}
\label{Eq5}
\omega_0^2 > \sum_\mu \frac{m_\mu}{m}\omega^2_\mu\lambda^2_\mu .
\end{equation}

It is convenient to rewrite our Hamiltonian in terms of the familiar annihilation and creation operators:
\begin{eqnarray}
\label{Eq6}
\hat{a} = \sqrt{\frac{m\omega_0}{2\hbar}}\left(\hat{x} + \frac{i\hat{p}}{m\omega_0}\right)  \nonumber \\
\hat{b}_\mu = \sqrt{\frac{m_\mu\omega_\mu}{2\hbar}}\left(\hat{x}_\mu + \frac{i\hat{p}_\mu}{m_\mu\omega_\mu}\right)
\end{eqnarray}
In terms of these operators our Hamiltonian becomes
\begin{equation}
\label{Eq7}
\hat{H} = \hbar\omega_0\hat{a}^\dagger\hat{a} + \sum_\mu\hbar\omega_\mu\hat{b}_\mu^\dagger\hat{b}_\mu
+ \sum_\mu\frac{\hbar}{2} V_\mu\left(\hat{a} + \hat{a}^\dagger\right)\left(\hat{b}_\mu + \hat{b}^\dagger_\mu\right)
\end{equation}
when unimportant constant shifts in the ground-state energies are removed and
\begin{equation}
\label{Eq8}
V_\mu = - \sqrt{\frac{m_\mu\omega_\mu}{m\omega_0}}\omega_\mu\lambda_\mu .
\end{equation}
When written in terms of this quantity, our positivity condition (\ref{Eq5}) becomes
\begin{equation}
\label{Eq9}
\omega_0 > \sum_\mu \frac{V^2_\mu}{\omega_\mu} .
\end{equation}

There is one further refinement that we need to introduce before considering the diagonalisation of
the Hamiltonian and this it to introduce a continuum description of the environment.  To this end we
introduce continuum annihilation and creation operators, $\hat{b}(\omega)$ and $\hat{b}^\dagger(\omega)$,
satisfying the commutation relations
\begin{equation}
\label{Eq10}
\left[\hat{b}(\omega), \hat{b}^\dagger(\omega')\right] = \delta(\omega - \omega') 
\end{equation}
and our Hamiltonian becomes \cite{Huttner1992b}
\begin{eqnarray}
\label{Eq11}
\hat{H} = \hbar\omega_0\hat{a}^\dagger\hat{a} + \int_0^\infty d\omega \:\hbar\omega \hat{b}^\dagger(\omega)\hat{b}(\omega) \nonumber \\
  \qquad \qquad +\int_0^\infty d\omega \: \frac{\hbar}{2}V(\omega)\left(\hat{a} + \hat{a}^\dagger\right)\left[\hat{b}^\dagger(\omega)
+\hat{b}(\omega)\right] 
\end{eqnarray}
and the positivity condition is
\begin{equation}
\label{Eq12}
\omega_0 > \int_0^\infty d\omega \frac{V^2(\omega)}{\omega} .
\end{equation}
Our Hamiltonian is quadratic in the annihilation and creation operators for the oscillator and the reservoir
and hence leads to linear coupled equations of motion for these operators.  We could seek to solve these
equations of motion and this would lead to an operator Heisenberg-Langevin equation analogous
to (\ref{Eq1}).  Here we adopt the different approach of diagonalising the Hamiltonian.

\subsection{Hamiltonian diagonalisation}
We seek to diagonalise the Hamiltonian by finding a complete set of eigenoperators, $\hat{B}(\omega)$
and their conjugates $\hat{B}^\dagger(\omega)$ that satisfy the operator equations
\begin{eqnarray}
\label{Eq13}
\left[\hat{B}(\omega), \hat{H}\right] = \hbar\omega \hat{B}(\omega) \nonumber \\
\left[\hat{B}^\dagger(\omega), \hat{H}\right] = -\hbar\omega \hat{B}^\dagger(\omega)
\end{eqnarray}
for all positive frequencies $\omega$.  These eigenoperators are complete if, in addition to these 
they also satisfy the condition
\begin{equation}
\label{Eq14}
\left[\hat{B}(\omega),\hat{B}^\dagger(\omega')\right] = \delta(\omega-\omega')  .
\end{equation}
These operator equations are the natural analogues of the more familiar eigenvalue and completeness
conditions for the eigenstates of a Hamiltonian.  In analogy with the eigenvalue problem, we expand each 
of the eigenoperators as a superposition of a complete set of operators:
\begin{equation}
\label{Eq15}
\hat{B}(\omega) = \alpha(\omega)\hat{a} + \beta(\omega)\hat{a}^\dagger + 
\int_0^\infty d\omega' \: \left[\gamma(\omega,\omega')\hat{b}(\omega') + \delta(\omega,\omega')\hat{b}^\dagger(\omega')\right] ,
\end{equation}
and then use the eigenoperator equations and completeness condition to determine the coefficients in
this expansion.  The calculation is a little involved but the main points are summarised in \ref{Fano}.

We can express any of the annihilation and creation operators for the oscillator or the reservoir in terms
of our eigenoperators.  To do this we write the desired operator as a superposition of all the $\hat{B}(\omega)$
and $\hat{B}^\dagger(\omega)$ operators and then use the commutation relations to extract the coefficients 
in this expansion.  For the oscillator operators we find
\begin{eqnarray}
\label{Eq16}
\hat{a} = \int_0^\infty d\omega \left(\alpha^*(\omega)\hat{B}(\omega) - \beta(\omega)\hat{B}^\dagger(\omega)\right)
\nonumber \\
\hat{a}^\dagger = \int_0^\infty d\omega \left(\alpha(\omega)\hat{B}^\dagger(\omega) - \beta^*(\omega)\hat{B}(\omega)\right) .
\end{eqnarray}
The requirement that these operators satisfy the familiar boson commutation relation, 
$\left[\hat{a},\hat{a}^\dagger\right] = 1$, provides a constraint on the functions $\alpha(\omega)$ and $\beta(\omega)$
in the form
\begin{equation}
\label{Eq17}
\int_0^\infty d\omega \left[|\alpha(\omega)|^2 - |\beta(\omega)|^2\right] = 
\int_0^\infty d\omega |\alpha(\omega)|^2\frac{4\omega_0\omega}{(\omega_0 + \omega)^2} = 1 ,
\end{equation}
where we have used (\ref{EqA10}).  It is interesting to note that the correctness of this may be verified
explicitly by contour integration, but that the proof makes explicit use of the positivity condition
(\ref{Eq12}) \cite{Huttner1992b}.

The integrand in (\ref{Eq17}) is clearly positive (or zero) for all frequencies and hence has the mathematical
form of a frequency probability distribution:
\begin{equation}
\label{Eq18}
\pi(\omega) =  |\alpha(\omega)|^2\frac{4\omega_0\omega}{(\omega_0 + \omega)^2}  .
\end{equation} 
A number of further constraints on this quantity emerge naturally from thinking of it as a probability 
distribution and from our diagonalisation. 

\subsection{Physical constraints}

We leave until the next section the question of the physical interpretation of the probability density
$\pi(\omega)$ but consider here what can be inferred from the fact that it has the mathematical 
properties of a probability distribution.  To this end let us denote the average value of a function
of frequency for this distribution by
\begin{equation}
\label{Eq19}
\langle\langle f(\omega) \rangle\rangle = \int_0^\infty d\omega f(\omega)\pi(\omega) .
\end{equation}
we note that $|\alpha(\omega)|^2$ is finite for all $\omega$ and it follows, therefore,
from (\ref{Eq18}) that the average value $\langle\langle\omega^{-1}\rangle\rangle$ is finite.

It follows from the eigenoperator equations (\ref{Eq13}) that we can write the Hamiltonian in the
form
\begin{equation}
\label{Eq20}
\hat{H} = \int_0^\infty d\omega \:\hbar\omega\: \hat{B}^\dagger(\omega)\hat{B}(\omega) + C  ,
\end{equation}
where $C$ is an unimportant constant.  If we substitute into this Hamiltonian our expression for
the eigenoperators (\ref{Eq15}) we require that the coefficients of $\hat{a}^2$ and $\hat{a}^{\dagger2}$
should vanish so that
\begin{equation}
\label{Eq21}
\int_0^\infty d\omega \:\omega |\alpha(\omega)|^2\left(\frac{\omega - \omega_0}{\omega + \omega_0}\right) = 
\int_0^\infty d\omega \frac{\pi(\omega)}{4\omega_0}(\omega^2 - \omega_0^2)
= 0 
\end{equation}
which implies that 
\begin{equation}
\label{Eq22}
\langle\langle \omega^2 \rangle \rangle = \omega^2_0 .
\end{equation}
The fact that the square of the mean value cannot exceed the mean value of the square for any
probability distribution leads us to deduce that
\begin{equation}
\label{Eq23}
\langle\langle\omega\rangle\rangle < \omega_0 .
\end{equation}

Finally we can apply the Cauchy-Schwartz inequality to provide a lower bound on the value of 
$\langle\langle\omega^{-1}\rangle\rangle$:
\begin{eqnarray}
\label{Eq24}
\langle\langle\omega^{-1}\rangle\rangle \langle\langle\omega\rangle\rangle \ge  1 
\nonumber \\
\Rightarrow \quad \langle\langle\omega^{-1}\rangle\rangle > \frac{1}{\omega_0} .
\end{eqnarray}
These inequalities are useful in determining the properties of the ground state.


\section{Ground-state}
It is immediately clear from the form of the Hamiltonian (\ref{Eq11}) that the ground state of the oscillator
is not that of the undamped oscillator, that is the state annihilated by $\hat{a}$.  To see this we need
only note that the interaction term $\hat{a}^\dagger\hat{b}^\dagger(\omega)$ acts on the combined ground 
state of the non-interacting oscillator and reservoir, adding a quantum to each.  This suggests that the 
true ground state should be a superposition of states with varying numbers of quanta in both the oscillator
and the reservoir and hence an entangled state.  This situation is reminiscent of the ground state of
an atom in quantum electrodynamics, which is dressed by virtual photons \cite{Franco}.  The dressing
of the ground-state atom is responsible for a number of important effects including the Casimir-Polder
interaction \cite{Casimir,Edwin} and the form of the polarizability of the atom \cite{Rodney2006,Milonni2008}.
It is reasonable to expect that it will be similarly significant for our strongly damped oscillator.

The true ground state, which we denote by the ket $|0\rangle$, is the zero-eigenvalue eigenstate of
all the annihilation operators $\hat{B}(\omega)$:
\begin{equation}
\label{Eq25}
\hat{B}(\omega)|0\rangle = 0 \qquad \qquad \forall \: \omega .
\end{equation} 
The ground-state properties of the oscillator in this pure state are described by a mixed state density
operator obtained by tracing out the environmental degrees of freedom:
\begin{equation}
\label{Eq26}
\hat{\rho}_{\rm Osc} = {\rm Tr}_{\rm Env}\left(|0\rangle\langle 0|\right) .
\end{equation}
The most straightforward way to determine the form of this mixed state is to use the characteristic
function \cite{Radmore1997}:
\begin{eqnarray}
\label{Eq27}
\chi(\xi) &=& {\rm Tr}\left[\hat{\rho}\exp\left(\xi\hat{a}^\dagger - \xi^*\hat{a}\right)\right] \nonumber \\
&=& \langle 0|\exp\left(\xi\hat{a}^\dagger - \xi^*\hat{a}\right)|0\rangle .
\end{eqnarray}
This function provides a complete description of the state and all of its statistical properties.  To evaluate 
it we express $\hat{a}$ and $\hat{a}^\dagger$ in terms of the eigenoperators\footnote{We make 
use of the operator ordering theorem \cite{Radmore1997}
\begin{eqnarray}
\nonumber
\exp(\hat{A}+\hat{B}) = \exp(\hat{A})\exp(\hat{B})
\exp\left(-\frac{1}{2}[\hat{A},\hat{B}]\right) ,
\end{eqnarray}
which holds if the two operators $\hat{A}$ and $\hat{B}$ both commute with their commutator 
$[\hat{A},\hat{B}]$.}
\begin{eqnarray}
\label{Eq28}
\chi(\xi) &=& \langle 0|\exp\left\{\int_0^\infty  d\omega
\left[\left(\xi\alpha+\xi^*\beta\right)\hat{B}^\dagger(\omega)
- \left(\xi^*\alpha^*+\xi\beta^*\right)\hat{B}(\omega)
\right]\right\}|0\rangle \nonumber \\
&=& \exp\left(-\frac{1}{2}\int_0^\infty d\omega|\xi\alpha(\omega)+\xi^*\beta(\omega)|^2\right) .
\end{eqnarray}
This simple form, Gaussian in $\xi$, is characteristic of a squeezed thermal state.  We can 
rewrite this characteristic function in terms of our probability density in the form
\begin{eqnarray}
\label{Eq29}
\chi(\xi) &=& \exp\left[-\frac{1}{2}\int_0^\infty \pi(\omega)\left(\frac{\omega}{\omega_0}\xi_r^2
+ \frac{\omega_0}{\omega}\xi_i^2\right)\right] \nonumber \\
&=& \exp\left[-\frac{1}{2}\left(\frac{\langle\langle\omega\rangle\rangle}{\omega_0}\xi_r^2
+ \langle\langle\omega^{-1}\rangle\rangle\omega_0\:\xi_i^2\right)\right] ,
\end{eqnarray}
where $\xi_r$ and $\xi_i$ are the real and imaginary parts of $\xi$ respectively.  We note, in
passing, that the quadrature operators for the oscillator (familiar from quantum optics \cite{Radmore1997})
have unequal uncertainties:
\begin{eqnarray}
\label{Eq30}
\Delta\left(\frac{\hat{a}+\hat{a}^\dagger}{\sqrt{2}}\right) = \sqrt{\frac{\langle\langle\omega^{-1}\rangle\rangle\omega_0}{2}}
\nonumber \\
\Delta\left(\frac{-i(\hat{a}-\hat{a}^\dagger)}{\sqrt{2}}\right) = 
\sqrt{\frac{\langle\langle\omega\rangle\rangle}{2\:\omega_0}}  .
\end{eqnarray}
The product of these two exceeds $\frac{1}{2}$, as it must, by virtue of the Cauchy-Schwartz inequality (\ref{Eq24}).

We have established that in the ground state, the reduced state of the oscillator is Gaussian in position and
momentum and, as such, is completely determined by the first and second moments of the position and
momentum.  These moments are
\begin{eqnarray}
\label{Eq31}
\langle 0|\hat{x}|0\rangle = 0 \nonumber \\
\langle 0|\hat{p}|0\rangle = 0 \nonumber \\
\langle 0|\hat{x}^2|0\rangle = \frac{\hbar\langle\langle\omega^{-1}\rangle\rangle}{2m} \nonumber \\
\langle 0|\hat{p}^2|0\rangle = \frac{\hbar m\langle\langle\omega\rangle\rangle}{2} \nonumber \\
\langle 0|\hat{p}\hat{x} + \hat{p}\hat{x}|0\rangle = 0 .
\end{eqnarray}
We note that the frequency of the undamped oscillator, $\omega_0$, makes no explicit appearance in these 
expectation values.  The mean energy of the oscillator alone is 
\begin{equation}
\label{Eq32}
\langle0|\frac{\hat{p}^2}{2m} + \frac{1}{2}m\omega^2_0\hat{x}^2|0\rangle =
\frac{1}{4}\hbar\omega_0\left(\frac{\langle\langle\omega\rangle\rangle}{\omega_0} + 
\langle\langle \omega^{-1}\rangle\rangle\omega_0\right) > \frac{1}{2}\hbar\omega_0 .
\end{equation}
The fact that this exceeds the ground-state energy of the undamped oscillator is a reflection of the 
fact that there is an energy cost to be paid in order to decouple the oscillator from its environment
\cite{Allahverdyan}.  We note, further, that the mean kinetic energy and potential energy for the oscillator 
alone do not have the same value and that this is in marked contrast to the ground state of the 
undamped oscillator.  

We can diagonalise the density operator for the oscillator alone, $\hat{\rho}_{\rm Osc}$, 
by means of a squeezing transformation
\cite{Radmore1997} or, equivalently, introducing a new pair of annihilation and creation operators
corresponding to a new oscillation frequency, $\omega_c$:
\begin{eqnarray}
\label{Eq33}
\hat{c} = \sqrt{\frac{m\omega_c}{2\hbar}}\left(\hat{x} + i\frac{\hat{p}}{m\omega_c}\right) \nonumber \\
\hat{c}^\dagger = \sqrt{\frac{m\omega_c}{2\hbar}}\left(\hat{x} - i\frac{\hat{p}}{m\omega_c}\right) .
\end{eqnarray}
To complete the diagonalisation we need only choose $\omega_c$ such that the expectation values
of $\hat{c}^2$ and $\hat{c}^{\dagger 2}$ are zero:
\begin{eqnarray}
\label{Eq34}
\langle 0|\hat{c}^2|0\rangle = \frac{m\omega_c}{2\hbar}
\langle 0|\hat{x}^2 - \frac{\hat{p}^2}{m^2\omega_c^2}|0\rangle  = 0 \nonumber \\
\Rightarrow \omega_c = \sqrt{\frac{\langle\langle\omega\rangle\rangle}{\langle\langle\omega^{-1}\rangle\rangle}} .
\end{eqnarray}
This frequency is, by virtue of (\ref{Eq23}) and (\ref{Eq24}), less than $\omega_0$, the natural frequency
of the undamped oscillator.  The appearance of a reduced oscillation frequency is reminiscent of the 
observed reduction in the oscillation frequency of a damped classical oscillator but unlike in the classical
case, the frequency $\omega_c$ remains real even in the heavily damped regime.  We note that Philbin and 
Horsley arrived at a similar reduced but always real frequency, which they associated with a decrease in 
the zero point energy \cite{Philbin2013}.

The mean number of $c$-quanta in the oscillator ground-state is
\begin{eqnarray}
\label{Eq35}
\bar{n}_c = \langle 0|\hat{c}^\dagger\hat{c}|0\rangle = 
\frac{1}{2}\left(\sqrt{\langle\langle\omega\rangle\rangle\langle\langle\omega^{-1}\rangle\rangle} - 1\right) ,
\end{eqnarray}
which exceeds $0$, as it should, by virtue of (\ref{Eq24}).  When written in terms of the $c$-quanta, the
steady-state density operator takes the form of a thermal Bose-Einstein state, which we can write in the
form
\begin{equation}
\label{Eq36}
\hat{\rho}_{\rm Osc} = \frac{1}{\bar{n}_c + 1}\left(\frac{\bar{n}_c}{\bar{n}_c + 1}\right)^{\hat{c}^\dagger\hat{c}} .
\end{equation}
We may interpret this state as a thermal state for the oscillator at the shifted (and reduced) frequency
$\omega_c$ and at an effective ``temperature"
\begin{equation}
\label{Eq37}
T_{\rm eff} = \frac{\hbar\omega_c}{k_B\ln(1 + \bar{n}_c^{-1})} .
\end{equation}
We should note, however, that the true temperature in the ground-state is zero and that this quantity
and the frequency $\omega_c$ are at most only parameters with which to quantify the state of the
oscillator and its entanglement with the environment.  In particular, the von-Neumann entropy associated 
with the steady state of the oscillator is
\begin{equation}
\label{Eq38}
S({\rm Osc}) = (\bar{n}_c + 1)\ln(\bar{n}_c + 1) - \bar{n}_c \ln\bar{n}_c ,
\end{equation}
and that, by virtue of the Araki-Lieb inequality \cite{Araki,Wehrl,QIbook} and the fact that the full state is pure, 
means that this is also the total entropy of the environment:
\begin{equation}
\label{Eq39}
S({\rm Env}) = S({\rm Osc}) 
\end{equation}
and that the quantum mutual information \cite{QIbook}, or index of correlation \cite{Simon}, between the
oscillator and the environment is
\begin{equation}
\label{Eq40}
S({\rm Osc}:{\rm Env}) = S({\rm Env}) + S({\rm Osc}) - S({\rm Osc},{\rm Env}) = 2S({\rm Osc}) .
\end{equation}


\section{Physical meaning of $\pi(\omega)$}

We have seen that the physical properties of the oscillator ground state may readily be expressed in
terms of moments of the frequency given the probability distribution $\pi(\omega)$.  Here we present 
the case for identifying this probability density with the contribution from the dressed modes, associated
with the eigenoperators $\hat{B}(\omega)$, to the the state of the oscillator.  It is in, in essence, the spectrum
of the true ground-state modes contributing to the oscillator state.  We present three 
arguments to support this interpretation.

Our first justification arises from the form of the expectation values (\ref{Eq31}).  We know that the
ground state of a harmonic oscillator of frequency $\omega$ has 
\begin{eqnarray}
\label{Eq41}
\langle\hat{x}^2\rangle = \frac{\hbar}{2m\omega} \nonumber \\
\langle\hat{p}^2\rangle = \frac{\hbar m\omega}{2} .
\end{eqnarray}
If we treat the state of the oscillator as a mixture of oscillators of different frequencies, each in its ground state,
and contribution with weight $\pi(\omega)$ then the resulting average mean-square values will be
\begin{eqnarray}
\label{Eq42}
\langle\hat{x}^2\rangle = \int_0^\infty d\omega \: \pi(\omega) \frac{\hbar}{2m\omega} 
= \frac{\hbar\langle\langle\omega^{-1}\rangle\rangle}{2m}\nonumber \\
\langle\hat{p}^2\rangle = \int_0^\infty d\omega \: \pi(\omega)\frac{\hbar m\omega}{2} 
= \frac{\hbar m\langle\langle\omega\rangle\rangle}{2} ,
\end{eqnarray}
which correspond to those obtained above.

Our second point arises from the form of the Hamiltonian for the oscillator
\begin{equation}
\label{Eq43}
\hat{H}_0 = \frac{\hat{p}^2}{2m} + \frac{1}{2}m\omega^2_0\hat{x}^2 .
\end{equation}
We can, by virtue of (\ref{Eq22}), write this as a combination of potentials corresponding to different 
frequencies but weighted by $\pi(\omega)$:
\begin{eqnarray}
\label{Eq44}
\hat{H}_0 &=& \frac{\hat{p}^2}{2m} + \frac{1}{2}m\int_0^\infty d\omega \: \pi(\omega) \omega^2 \hat{x}^2
\nonumber \\
&=& \frac{\hat{p}^2}{2m} + \frac{1}{2}m \langle\langle\omega^2\rangle\rangle \hat{x}^2 .
\end{eqnarray}

Finally we note that the effective mean energy of the oscillator, which is associated with the diagonal
form of the density operator (\ref{Eq36}) is
\begin{equation}
\label{Eq45}
\left(\bar{n}_c + \frac{1}{2}\right)\hbar\omega_c = \frac{1}{2}\hbar\int_0^\infty d\omega \: \pi(\omega) \omega
= \frac{1}{2}\hbar\langle\langle\omega\rangle\rangle ,
\end{equation}
which combines the characteristic ground state energies of the dressed oscillators, weighted by the 
probability distribution $\pi(\omega)$.  The combination of these three features leads us to interpret
$\pi(\omega)$ as the proportion of the corresponding dressed oscillators contributing to the properties
of the damped oscillator.  We emphasise that the mathematical results obtained in the preceding section 
do not \emph{require} us to adopt this interpretation of $\pi(\omega)$ but we find it helpful to do so.


\section{Oscillator dynamics}

The exact diagonalisation of the Hamiltonian makes it straightforward to evaluate the time-evolution of
any desired property of the oscillator.  All that we need do is to express the desired observable in terms
of the eigenoperators, $\hat{B}(\omega)$ and $\hat{B}^\dagger(\omega)$ and then use the time evolution
of these operators, the form of which is an elementary consequence of the fact that they are eigenoperators:
\begin{eqnarray}
\label{Eq46}
\hat{B}(\omega,t) = \hat{B}(\omega,0)e^{-i\omega t} \nonumber \\
\hat{B^\dagger}(\omega,t) = \hat{B}^\dagger(\omega,0)e^{i\omega t} .
\end{eqnarray}
In particular, we can determine the time-evolution of the annihilation operator for the oscillator in this 
way:
\begin{eqnarray}
\label{Eq47}
\hat{a}(t) &=& \int_0^\infty d\omega \left[\alpha^*(\omega)\hat{B}(\omega,0)e^{-i\omega t}
-\beta(\omega)\hat{B}^\dagger(\omega,0)e^{i\omega t}\right]  \nonumber \\
&=& \left.  \int_0^\infty d\omega \right\{\alpha^*(\omega)e^{-i\omega t}
\left[\alpha(\omega)\hat{a}(0) + \beta(\omega)\hat{a}^\dagger(0) +  \right. \nonumber \\
& & \qquad  \int_0^\infty d\omega' \left. \: \left(\gamma(\omega,\omega')\hat{b}(\omega',0) + 
\delta(\omega,\omega')\hat{b}^\dagger(\omega',0)\right) \right] \nonumber \\
& & \qquad -\beta(\omega)e^{i\omega t}\left[\alpha^*(\omega)\hat{a}^\dagger(0) + \beta^*(\omega)\hat{a}(0) + \right. 
\nonumber \\
& & \qquad  \left. \int_0^\infty d\omega' \left.\: \left(\gamma^*(\omega,\omega')\hat{b}^\dagger(\omega',0) + 
\delta^*(\omega,\omega')\hat{b}(\omega',0)\right) \right] \right\} .
\end{eqnarray}
This may be used, together with the initial state of the oscillator and the environment, to evaluate the 
expectation value of any desired property.  We note that if the environment is in a stationary state, so 
that $\langle\hat{b}(\omega,0)\rangle = 0 = \langle\hat{b}^\dagger(\omega,0)\rangle $, then the expectation
values of the position and momentum operators take a pleasingly simple form:
\begin{eqnarray}
\label{Eq48}
\langle\hat{x}(t)\rangle = \langle\langle\cos(\omega t)\rangle\rangle \langle\hat{x}(0)\rangle 
+ \frac{1}{m}\langle\langle\omega^{-1}\sin(\omega t)\rangle\rangle \langle\hat{p}(0)\rangle \nonumber \\
\langle\hat{p}(t)\rangle = \langle\langle\cos(\omega t)\rangle\rangle \langle\hat{p}(0)\rangle 
- m\langle\langle\omega\sin(\omega t)\rangle\rangle \langle\hat{x}(0)\rangle ,
\end{eqnarray}
where the double angle brackets denote, as before, averages over our probability distribution $\pi(\omega)$
as in (\ref{Eq19}).  The form of these equations adds further support to our interpretation of $\pi(\omega)$
as a frequency probability distribution for the damped oscillator, as they may be viewed as the evolution
of an undamped oscillator with a frequency $\omega$ averaged using this probability distribution.  The
dissipation arises simply from a dephasing amongst the different frequency components.

The evolution of the mean position and momentum, as given in (\ref{Eq48}), has the necessary 
short-time form of that for an undamped oscillator
\begin{eqnarray}
\label{Eq49}
\langle\hat{x}(dt)\rangle = \langle\hat{x}(0)\rangle + \frac{dt}{m}\langle\hat{p}(0)\rangle \nonumber \\
\langle\hat{p}(dt)\rangle = \langle\hat{p}(0)\rangle - m\omega_0^2dt\langle\hat{x}(0)\rangle ,
\end{eqnarray}
where we have used the identity $\langle\langle\omega^2\rangle\rangle = \omega_0^2$.  The effects
of the coupling to the environment enter at order $dt^3$ and this is an indication of the essentially
non-Markovian nature of the strongly-damped oscillator.  Our interest
is in strongly damped oscillators and so we should note that (\ref{Eq48}) includes the possibilities of
both critically-damped and over-damped evolution.  The equations contain, moreover, a simple criterion
for these, which we may express in terms of our probability distribution.  The motion will be oscillatory
if $\langle\langle\cos(\omega t)\rangle\rangle$ has stationary points at times other than at $t=0$.  Alternatively,
we may state that the motion is under-damped if the derivative of this quantity, that is 
$\langle\langle\omega\sin(\omega t)\rangle\rangle$, is zero for any time other than $t=0$.  If it is zero
only at $t=0$ then the motion is critically-damped or over-damped.

Our expression for the evolved annihilation operator (\ref{Eq47}), together with the corresponding one
for the creation operator provide a full description of the oscillator dynamics.  This is true for any initial
state of the oscillator and, moreover, for any environmental state including, of course, that associated with
a finite temperature.  As an illustration let us examine the evolution of the characteristic function for
an arbitrary initial state of the oscillator coupled to a zero-temperature environment at time $t = 0$.  
With a little effort we find
\begin{equation}
\label{Eq50}
\chi(\xi,t) = \chi[\xi(t),0]\exp\left[-\frac{1}{2}\int_0^\infty d\omega' 
\left|\int_0^\infty d\omega \: \mu(\omega,\omega',t)\right|^2\right] ,
\end{equation}
where
\begin{eqnarray}
\label{Eq51}
\xi(t) = \int_0^\infty d\omega \left[\eta(\omega,t)\alpha^*(\omega) - \eta^*(\omega,t)\beta(\omega)\right]
\nonumber \\
\mu(\omega,\omega',t) = \eta(\omega,t)\gamma^*(\omega,\omega') - \eta^*(\omega,t)\delta^*(\omega,\omega')
\nonumber \\
\eta(\omega,t) = \left[\xi\alpha(\omega)+\xi^*\beta(\omega)\right]e^{i\omega t} .
\end{eqnarray}
This characteristic function encodes the full dynamics and statistics of the oscillator.  As a simple illustration
of this we can determine the form of the steady state.  To see this we first note $\xi(t)$ tends to zero as
$t$ tends to infinity and the different frequency components dephase so that
\begin{equation}
\label{Eq52}
\chi[\xi(\infty),0] = \chi[0,0] = 1 .
\end{equation}
Evaluating the long-time limit of the exponential factor in (\ref{Eq50}) requires some care in the handling of 
the delta-function and principal part components.  We find
\begin{equation}
\label{Eq53}
\int_0^\infty d\omega' 
\left|\int_0^\infty d\omega \: \mu(\omega,\omega',\infty)\right|^2
 = \int_0^\infty d\omega \left|\xi\alpha(\omega') + \xi^*\beta(\omega')\right|^2 ,
\end{equation}
so the steady-state characteristic function is
\begin{equation}
\label{Eq54}
\chi(\xi,\infty) = \exp\left(-\frac{1}{2}\int_0^\infty d\omega\left|\xi\alpha(\omega) + \xi^*\beta(\omega)\right|^2\right) ,
\end{equation}
which we recognise as the characteristic function for the oscillator in the global ground state (\ref{Eq28}).
This is a most satisfactory and exact result.


\section{Weak-coupling limit}
The theory developed above was designed to treat the strongly-damped harmonic oscillator, but should
also be applicable to the more familiar weakly damped oscillator, for which the oft-employed Born and
Markov approximations are applicable and the steady state of the oscillator should be its ground state.
We show here that this is indeed the case.

We start by considering the form of the function $\alpha(\omega)$ in the weak coupling limit.  To aid
our analysis we rewrite the form given in (\ref{EqA18}) as
\begin{equation}
\label{Eq55}
\alpha(\omega) = \frac{\omega + \omega_0}{\omega_0}\left(\frac{V(\omega)}{|V(\omega)|^2Y(\omega)
-i\pi|V(\omega)|^2}\right) .
\end{equation}
The weak damping limit corresponds to choosing the coupling to the environment to be small or,
more specifically, to $|V(\omega)|^2 \ll \omega_0$.  It is clear that in this limit, $|\alpha(\omega)|^2$
will be a sharply peaked function centred around the frequency for which $Y(\omega) = 0$.  If the
integral part in $Y(\omega)$, as given in (\ref{EqA15}) is small\footnote{If it is not then we will need
to invoke ideas of renormalisation, a complication we wish to avoid.} then this frequency will be close
to $\omega_0$ and we can write
\begin{equation}
\label{Eq56}
|V(\omega)|^2Y(\omega) \approx 4(\omega - \omega_0) - 4F(\omega) ,
\end{equation}
where 
\begin{equation}
\label{Eq57}
4F(\omega) = \int_0^\infty d\omega' 
\left(\frac{\mathbbmss{P}}{\omega-\omega'} - \frac{1}{\omega+\omega'}\right)|V(\omega')|^2 .
\end{equation}
This leads, in turn, to a corresponding approximate form for $\alpha(\omega)$:
\begin{equation}
\label{Eq58}
\alpha(\omega) \approx \frac{V(\omega)}{2}\frac{1}{\omega - \omega_0 - F(\omega) - i\frac{\pi}{4}|V(\omega)|^2} ,
\end{equation}
where we have set $\omega = \omega_0$ everywhere except in the denominator.  We note that this is of 
the form that arises from the Fano diagonalisation of our problem if we make the rotating wave approximation 
by omitting from our original Hamiltonian all terms that are products of two creation operators or of two annihilation
operarors \cite{Radmore1997,Radmore1988}.

Consistency with the above approximation, which led us to set $\omega = \omega_0$ leads us to 
set $\beta(\omega)$ to zero:
\begin{equation}
\label{Eq59}
\beta(\omega) \approx 0 ,
\end{equation}
so that the integral over all frequency of $|\alpha(\omega)|^2$ is unity.  Hence in this limit the steady-state
characteristic function for our oscillator is
\begin{equation}
\chi(\xi,\infty) = \exp\left(-\frac{1}{2}|\xi|^2\right) ,
\end{equation}
which we recognise as that for the ground state of the undamped oscillator \cite{Radmore1997}.
Further, we note that in this limit our probability distribution function, $\pi(\omega) \approx 
|\alpha(\omega)|^2$, approaches a Lorentzian centred on $\omega_0$, with some width $\gamma$
\footnote{This is not strictly true in the wings of the distribution, of course, as even in this limit we require 
$\langle\langle\omega^2\rangle\rangle = \omega_0^2$, but the corresponding quantity for a true 
Lorentzian is divergent.}.  Thus all the complexity of of the original problem is reduced, in the weak-damping 
limit to just two parameters: a natural oscillation frequency and a damping rate. 

It was important to confirm that our more general treatment coincided, in the right limit, with the 
approximate methods used for weakly damped oscillators.  We should note that even if we are
working in the weakly damped regime, then our approach offers a systematic way to treat corrections
to the results obtained using the Born and Markov approximations, which may play an important role
in modelling measurements at the limits of sensitivity.


\section{Conclusion}

We have presented an exact diagonalisation of a simple quantum model of the damped harmonic oscillator, 
one that is applicable, in particular, to strong damping.  We have found that much of the behaviour of the 
oscillator and may of its properties can be described in terms of a single probability function, $\pi(\omega)$,
which we may interpret as the contribution of corresponding dressed mode, at frequency $\omega$ to the
oscillator.  These properties include the steady state of the oscillator at zero temperature, the entanglement 
between the oscillator and its environment and also its evolution, both in the familiar under-damped regime 
but also in the more problematic over-damped regime.  

We have applied our diagonalisation to study the properties of the true ground state and have shown that 
the oscillator part of this pure entangled state coincides with the steady-state of the oscillator in a zero-temperature
environment.  The diagonalisation is not specific to any particular
state of the reservoir, however, and may readily be applied to environments at finite temperature.  It may be 
extended, moreover, to include driving forces, coupled oscillators and multiple reservoirs, with the latter perhaps 
being at different temperatures \cite{Martinez}.  This may provide some insights into important questions of 
principle in the nascent fields of quantum machines and quantum thermodynamics 
\cite{Gemmer,Joan2011,Joan2013,Roncaglia,Binder,Brandao}.  We return to these ideas elsewhere.


\ack
The Hamiltonian diagonalisation upon which much of this work is based was first calculated by Bruno Huttner,
in collaboration with SMB, 24 years ago in their work on the quantum electrodynamics of dielectric media.  We 
are most grateful to him and also to Paul Radmore for helpful comments and suggestions.


\appendix
\section{Fano diagonalisation}
\label{Fano}

We present a brief account of the exact diagonalisation of our Hamiltonian based on methods developed by Fano for the
study of configuration interactions \cite{Fano}.  This idea was applied extended to weakly-coupled oscillators in a quantum
study of damped cavity modes \cite{Radmore1997,Radmore1988}.  The extension to stronger couplings, with the inclusion of 
counter-rotating couplings, has been given before and applied to the quantum theory of light in dielectric media
\cite{Huttner1992a,Huttner1992b}.  We summarise here the analysis presented in \cite{Huttner1992b}.

Our task is to diagonalise the damped harmonic oscillator Hamiltonian\footnote{For the Hamiltonian of interest in this paper
the coupling, $V(\omega)$, is real but treating the problem with a more general complex coupling presents no additional
difficulties.}
\begin{eqnarray}
\label{EqA1}
\hat{H} = \hbar\omega_0\hat{a}^\dagger\hat{a} + \int_0^\infty d\omega \:\hbar\omega \hat{b}^\dagger(\omega)\hat{b}(\omega) \nonumber \\
  \qquad \qquad +\int_0^\infty d\omega \: \frac{\hbar}{2}\left(\hat{a} + \hat{a}^\dagger\right)\left[V(\omega)\hat{b}^\dagger(\omega)
+V^*(\omega)\hat{b}(\omega)\right] ,
\end{eqnarray}
by which we mean rewriting it in the form of a continuum of \emph{uncoupled} or dressed oscillators:
\begin{equation}
\label{EqA2}
\hat{H} = \int_0^\infty d\omega \:\hbar\omega\: \hat{B}^\dagger(\omega)\hat{B}(\omega) + C  ,
\end{equation}
where $C$ is an unimportant constant.  We proceed by writing the dressed annihilation operators, $\hat{B}(\omega)$,
as linear combinations of the bare operators for the oscillator and bath modes:
\begin{equation}
\label{EqA3}
\hat{B}(\omega) = \alpha(\omega)\hat{a} + \beta(\omega)\hat{a}^\dagger + 
\int_0^\infty d\omega' \: \left[\gamma(\omega,\omega')\hat{b}(\omega') + \delta(\omega,\omega')\hat{b}^\dagger(\omega')\right] ,
\end{equation}
where $\alpha{\omega}$, $\beta(\omega)$, $\gamma(\omega,\omega')$ and $\delta(\omega,\omega')$ are to be determined.

We require the operator $\hat{B}(\omega)$ to be associated with an uncoupled or dressed oscillator of angular frequency 
$\omega$.  This requires us to find its form such that the following pair of operator equations are satisfied for every frequency,
$\omega$:
\begin{eqnarray}
\label{EqA4}
\left[\hat{B}(\omega), \hat{H}\right] = \hbar\omega \hat{B}(\omega) \\
\label{EqA5}
\left[\hat{B}(\omega),\hat{B}^\dagger(\omega')\right] = \delta(\omega-\omega')  .
\end{eqnarray}
Substituting the ansatz (\ref{EqA3}) into (\ref{EqA4}) and comparing coefficients of the bare creation and annihilation 
operators leads to the set of couple equations:
\begin{eqnarray}
\label{EqA6}
\alpha(\omega)\omega_0 + \frac{1}{2}\int_0^\infty d\omega' \left[\gamma(\omega,\omega')V(\omega')
- \delta(\omega,\omega')V^*(\omega')\right] = \alpha(\omega)\omega \\
\label{EqA7}
-\beta(\omega)\omega_0 + \frac{1}{2}\int_0^\infty d\omega' \left[\gamma(\omega,\omega')V(\omega')
- \delta(\omega,\omega')V^*(\omega')\right] = \beta(\omega)\omega \\
\label{EqA8}
\frac{V^*(\omega')}{2}\left[\alpha(\omega) - \beta(\omega)\right] + \gamma(\omega,\omega')\omega'    = \gamma(\omega,\omega')\omega \\
\label{EqA9}
\frac{V(\omega')}{2}\left[\alpha(\omega) - \beta(\omega)\right] - \delta(\omega,\omega')\omega'    = \delta(\omega,\omega')\omega  .
\end{eqnarray}
Our method of solution is use these to determine the functions $\beta{\omega}$, $\gamma(\omega,\omega')$ and
$\delta(\omega,\omega')$ in terms of $\alpha{\omega}$ and then to determine this remaining function by enforcing
the commutation relation (\ref{EqA5}).  From (\ref{EqA6}) and (\ref{EqA7}) we see that
\begin{equation}
\label{EqA10}
\beta({\omega}) = \frac{\omega - \omega_0}{\omega + \omega_0}\: \alpha(\omega)  .
\end{equation}
If we use this to substitute for $\beta(\omega)$ into the remaining equations then we find
\begin{eqnarray}
\label{EqA11}
V^*(\omega')\frac{\omega_0}{\omega + \omega_0}\alpha(\omega) = \gamma(\omega,\omega')(\omega - \omega') \\
\label{EqA12}
V(\omega')\frac{\omega_0}{\omega + \omega_0}\alpha(\omega) = \delta(\omega,\omega')(\omega + \omega')  .
\end{eqnarray}
Solving the second of these presents no difficulty and we find
\begin{equation}
\label{EqA13}
\delta(\omega,\omega') = \left(\frac{1}{\omega + \omega'}\right)V(\omega')\frac{\omega_0}{\omega + \omega_0}\alpha(\omega) .
\end{equation}
The first, however, requires careful handling because the behaviour at $\omega = \omega'$.  Following
Fano \cite{Fano}, we adopt the method proposed by Dirac \cite{Dirac1927} and write
\begin{equation}
\label{EqA14}
\gamma(\omega,\omega') = \left(\frac{\mathbbmss{P}}{\omega - \omega'} + Y(\omega)\delta(\omega - \omega')\right)
V^*(\omega')\frac{\omega_0}{\omega + \omega_0}\alpha(\omega) ,
\end{equation}
where $\mathbbmss{P}$ denotes that the principal part is to be taken on integration and $Y(\omega)$ is a real
function, which we determine by substituting (\ref{EqA14}) into (\ref{EqA6}).  We find
\begin{eqnarray}
\label{EqA15}
Y(\omega) = \frac{1}{|V(\omega)|^2}\left[\frac{2(\omega^2-\omega_0^2)}{\omega_0} - \int_0^\infty d\omega' 
\left(\frac{\mathbbmss{P}}{\omega-\omega'} - \frac{1}{\omega+\omega'}\right)|V(\omega')|^2\right] . 
\nonumber  \\
\end{eqnarray}

If we substitute our operators, $\hat{B}(\omega)$, expressed in terms of the function $\alpha(\omega)$ into the
commutation relation (\ref{EqA6}) then we find
\begin{eqnarray}
\label{EqA16}
\left[\hat{B}(\omega),\hat{B}^\dagger(\omega')\right] = \alpha({\omega})\alpha^*(\omega')\left\{1 - 
\left(\frac{\omega-\omega_0}{\omega+\omega_0}\right)\left(\frac{\omega'-\omega_0}{\omega'+\omega_0}\right)\right.
\nonumber \\
 \quad + \int_0^\infty d\omega''
\left[\left(\frac{\mathbbmss{P}}{\omega-\omega''}+Y(\omega)\delta(\omega-\omega'')\right) 
\left(\frac{\mathbbmss{P}}{\omega'-\omega''}+Y(\omega)\delta(\omega'-\omega'')\right) \right.
\nonumber \\
\qquad \qquad \left. \left. - \left(\frac{1}{\omega + \omega''}\right) \left(\frac{1}{\omega' + \omega''}\right)\right]
\frac{|V(\omega'')|^2\omega_0^2}{(\omega+\omega_0)(\omega'+\omega_0)} \right\} .
\end{eqnarray}
Evaluating the integrals and setting the result equal to $\delta(\omega-\omega')$ gives\footnote{This requires the 
use of the following formula for the product of two principal parts \cite{Radmore1997}:
\begin{eqnarray}
\nonumber
\frac{\mathbbmss{P}}{\omega-\omega''}\frac{\mathbbmss{P}}{\omega'-\omega''} =
\frac{\mathbbmss{P}}{\omega'-\omega}\left(\frac{\mathbbmss{P}}{\omega-\omega''} -
\frac{\mathbbmss{P}}{\omega'-\omega''}\right) + \pi^2\delta(\omega-\omega'')\delta(\omega'-\omega'') .
\end{eqnarray}}
\begin{equation}
\label{EqA17}
|\alpha(\omega)|^2 = \frac{(\omega + \omega_0)^2}{\omega_0^2|V(\omega)|^2}
\left(\frac{1}{Y^2(\omega) + \pi^2}\right) .
\end{equation}
Note that the diagonalisation does not fix the phase of the complex function $\alpha(\omega)$ and we are 
free to choose this as we wish.  A convenient choice is to set
\begin{equation}
\label{EqA18}
\alpha(\omega) = \frac{\omega + \omega_0}{\omega_0V^*(\omega)}
\left(\frac{1}{Y(\omega) - i\pi}\right) .
\end{equation}


\end{document}